\documentclass[12pt]{article}
\usepackage{amssymb}
\usepackage{graphicx}
\usepackage{float}
\newcommand{\text}{\rm}

\newcommand{\bb}{\begin{equation}}
\newcommand{\ee}{\end{equation}}
\newcommand{\bega}{\begin{eqnarray}}
\newcommand{\ega}{\end{eqnarray}}
\newcommand{\begae}{\begin{eqnarray*}}
\newcommand{\egae}{\end{eqnarray*}}

\newcommand{\h}{\hspace*{4ex}}

\newcommand{\cent}{\centerline}
\newcommand{\vs}{\vspace*}

\begin{document}

\baselineskip 0.5cm

\begin{center}

{\large {\bf Generation and analysis of fractional Hankel-Bessel vortex beams via computational holography} }

\end{center}

\vs{0.2 cm}

\cent{George Brian dos Reis$^{\: 1}$, Rafael A. B. Suarez$^{\: 1}$, and Marcos R. R. Gesualdi$^{\: 1}$}

\vs{0.2 cm}

\centerline{{\em $^{\: 1}$ Universidade Federal do ABC, Santo Andr\'e, SP, Brazil.}}
\centerline{{\em $^{\: 2}$ Facultad de Ciencias Básicas, Universidad Santiago de Cali, Campus Pampalinda, Santiago de Cali,  Colombia.}}

\vs{0.5 cm}

{\bf Abstract  \ --} In this work, the experimental optical generation of the fractional Hankel-Bessel vortex beams were investigated through the holographic technique by means of computer generated holograms (CGHs) and reproduced in a spatial light modulator (SLMs).  The intensity and phase profiles were simulated and the experimental results are presented in this work, as well as their propagation along the z-axis is demonstrated. The experimental results are in accordance with the theoretical predictions described in the theory and literature. These results presents excellent perspectives of this optical vortex and potential applications in optical manipulation, optical microscopy and optics communications, optical metrology, among others.


\vs{0.5 cm}

\h {\em\bf 1. Introduction} 

An optical vortex (OV) is characterized by a helical phase structure that carries orbital angular momentum, whose magnitude depends on the topological charge $n$. This phase is superimposed onto the beams which lead to a phase singularity at the center of the radiation field, where the phase of the beams is undetermined and its amplitude vanishes \cite{allen1992orbital,zhang2022review,berry2004optical,shen2019optical}.

On the other hand, the non-diffracting waves or diffraction-resistant waves in optics are special optical beams that keep their intensity spatial shape during propagation. From the Helmholtz equation, a set of exact solutions characterize the non-diffracting waves described by a coordinate system in which the transverse and longitudinal components of this equation are separable, among which:  Cartesian, circular-cylindrical, elliptical-cylindrical and parabolic-cylindrical system; establishing plane waves, as well as the different types of beams, such as Bessel, Mathieu and Parabolic beam \cite{vieira2014modeling, gesualdi2023algoritmos2, suarez2016photorefractive, suarez2020generation}.

In addition, there are beams of structured light with orbital angular momentum (OAM), called optical vortices (OV) \cite{berry2004optical,shen2019optical}. These OVs have a phase peculiarity with a certain topological charge (TC), similar to a fluid vortex with a central flux singularity, which gives rise to a null intensity distribution and its amplitude disappears in the center of the beam. They are represented by a helical phase structure that carries OAM, and their magnitude depends on the TC \cite{yao2011orbital, zhang2019orbital, LopezMariscal2006}.

Its applications range from optical tweezers, optical metrology or applications in quantum information. Optical tweezers, in particular, can use OVs in the control and manipulation of micro-particles and nano-particles, since an OV can transfer OAM to these particles. This transfer occurs through the interaction between light and matter, resulting in an angular force \cite{zhang2022review, willner2015optical, yepes2019phase, suarez2020experimental, suarez2023optical, padgett2010optical}.

To understand the behavior of these OV along the propagation axis, it's necessary to apply optical and electronic techniques and instruments with high efficiency and measurement reliability. The holographic technique is a powerful optical beam generation technique that allows to generate and characterize this beams with high reliability and efficiency \cite{webster2006holographic, schnars2005hologram}.

Dennis Gabor, in 1948, was the first scientist to describe this technique in his work "A new microscopic principle" \cite{gabor1948}, when he observed that the interference of two waves contained information about the intensity (amplitude) and depth (phase) of the light scattered by an object and could be recorded on a holographic recording medium.

Holography is an interferometric-diffractive process, where the recording of an interference pattern produced by two waves on a "film" (holographic recording medium) is called a hologram. The reconstruction of the hologram is done by light diffracted in the grating recorded from the hologram, creating 3D images of the object. This hologram presents all depth perspective, since they carry intensity and phase information of the object in the reconstructed image \cite{hariharan1996, benton2008, cirino2011, gesualdi2023algoritmos, yepes2017dynamic}.

From the computational development of this technique, as well as spatial light modulators (SLM). Using computer generated holograms (CGH) \cite{Stuerwald2018}, this technique is capable of reproducing the behavior of the most diverse types of beams, diffractive or not, or of optical vortex, with very high precision and practicality \cite{arlt2001optical, coullet1989optical, stein1992, gesualdi2023algoritmos2, suarez2016photorefractive, suarez2020generation}.

Currently, Hankel-Bessel vortex beams are at the frontier of knowledge of these types of structured light with OAM, and their theoretical approach was described by Kotlyar at all \cite{kotlyar2012hankel}. Thus, this work deals with the experimental generation of this vortex using the holographic technique, in addition to considering a topological charge (TC) both integer and fractional, aiming at the different types of applications that fractional vortex have in optical manipulation techniques.

The study and propagation, theoretical and experimental, of these vortex are carried out from the solutions found in \cite{kotlyar2012hankel}. However, for the fractional vortex, it's necessary to employ an inverse Fourier transform. The results found are in line with what is predicted in the theory and literature considering the theoretical solutions presented and discussed by the authors in \cite{kotlyar2012hankel, kotlyar2018vortex, kovalev2013nonparaxial}. These results presents excellent perspectives of this optical vortex and potential applications in optical micro and nano manipulation \cite{suarez2020experimental, suarez2023optical}.

\h {\em\bf 2. Fractional Hankel-Bessel vortex beam} 

Many optical beams are found from the solutions of the Helmholtz equation. Among them, there are the paraxial hypergeometric beams (HyG beams) that have complex amplitude and are described by the Kummer function \cite{li2017oam}. These beams are found when calculating the integral for an angular spectrum of plane waves, considering that the spiral phase of the beam has even numbers in its topological charge, thus deriving a solution of the Helmholtz equation \cite{li2017oam, kotlyar2018vortex}.
\\
The product of two Kummer functions $_1F_1(a,b,x)$, describe the nonparaxial hypergeometric beams (nHyG beams) just by changing the value of x. These functions describe a superposition of two equivalent waves propagating in both the positive and negative direction of the z-axis \cite{kotlyar2018vortex, kovalev2013nonparaxial}. Assuming that in certain parameters of a function $U(a,b,x)$, proportional to the Hankel function and the Kummer function, is also proportional to the Bessel function, it's considered that the nonparaxial beams is called Hankel-Bessel (HB) vortex beam \cite{kotlyar2018vortex, kovalev2013nonparaxial}.
\\
Let's consider the following Helmholtz equation in the cylindrical coordinates $(r,\varphi,z)$ \cite{kotlyar2012hankel}.

\begin{equation}
\bigg(\frac{\partial^2}{\partial r^2}+\frac{1}{r}\frac{\partial}{\partial r}+\frac{1}{r^2}\frac{\partial}{\partial \varphi^2}+\frac{\partial^2}{\partial z^2}+k^2\bigg)E(r,\varphi,z)=0,
\label{eq:coord.cil}
\end{equation}
where $r$ is the position vector in the source plane, $k=2 \pi / \lambda$ is the wave number with the wavelength $\lambda$, $z$ is the propagation distance, and $\varphi$ is the azimuthal angle.

We are looking for a solution like $E(r,\varphi,z)=E(r,z)r^p exp(in\varphi+ikz)$, where $n$ and $p$ are integer numbers. Then, equation \ref{eq:coord.cil} becames:

\begin{equation}
\frac{\partial^2 E}{\partial r^2}+\frac{\partial^2 E}{\partial z^2}+\bigg(\frac{2p+1}{r}\bigg)\frac{\partial E}{\partial r}+2ik\frac{\partial E}{\partial z}+\bigg(\frac{p^2-n^2}{r^2}\bigg)E=0.
\label{eq:Eq.2}
\end{equation}

To avoid amplitude singularities, it's used $|p|$, so, at $n= \pm |p|$, the third term in equation \ref{eq:Eq.2} is eliminated. To solve this equation, let's changing to the parabolic coordinates, $u= \sqrt{r^2+z^2}+z^2$ e $v= \sqrt{r^2+z^2}-z^2$. Then, equation \ref{eq:Eq.2} takes the form.

\begin{equation}
u\frac{\partial^2 E}{\partial u^2} + v\frac{\partial^2 E}{\partial v^2} + (n+1+iku)\frac{\partial E}{\partial u}+(n+1-ikv)\frac{\partial E}{\partial v}=0.
\label{eq:Eq.3}
\end{equation}

\begin{figure}[H]
\centering
\includegraphics[width=13cm]{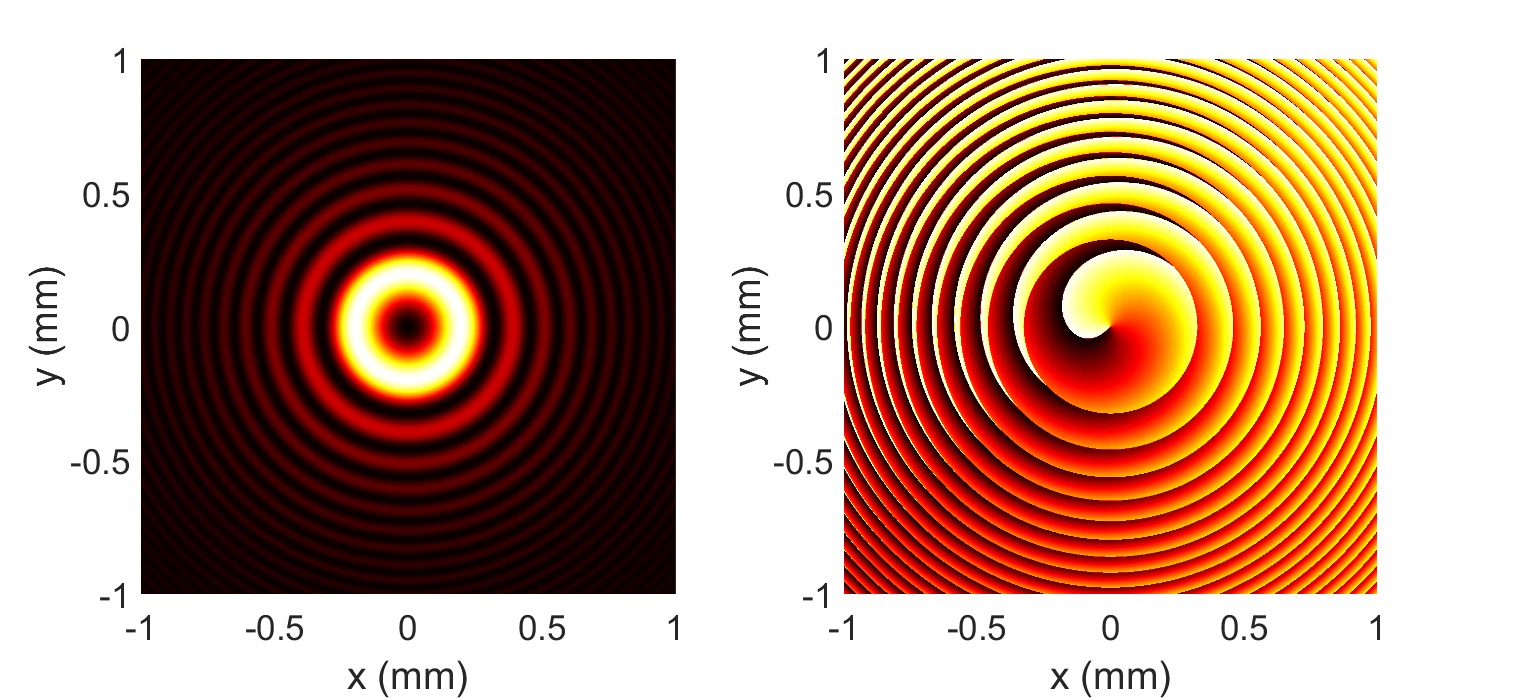}
\caption{Normalized intensity and phase map for the HB vortex beams with $n = 1$.}
\label{Fig_int_ph-1}
\end{figure}

This equation is solved in separable variables with their solutions described by Kummer's functions. After a series of mathematical deductions that we must describe the propagation of laser light in a certain direction, the relationship between Kummer’s function $_1F_1(a,b,z)$ and the second solution of Kummer’s equation, the solution of the original Helmholtz equation in \ref{eq:coord.cil},  can be written as \cite{kotlyar2012hankel, li2017oam}.:

\begin{equation}
E(r,\varphi,z)=i^{3n+1}\frac{\pi}{2} n! A_0 e^{(in\varphi)} \bigg[H_{\frac{n}{2}}^{(1)} \Omega_H\bigg] \bigg[J_{\frac{n}{2}}\Omega_J\bigg],
\label{eq:Sol_Eq_HB}
\end{equation}
where $n$ is the topological charge, $ A_0$ is a constant characterizing the beam power, $H_{\frac{n}{2}}^{(1)}$ is the first-order Hankel function, $\Omega_H = \frac{k}{2}\sqrt{r^2+z^2}+z$, $J_{\frac{n}{2}}$ is a Bessel function of integer and half-integer orders, and $\Omega_J = \frac{k}{2}\sqrt{r^2+z^2}-z$. 
\\
The HB vortex beam are described by Eq. \ref{eq:Sol_Eq_HB}. Furthermore, as they propagate along the positive z-axis, a divergence of the beams is observed and is proportional to $\sqrt{z}$. When we considered a large z, the HB vortex beam diverge hyperbolically.
\\
The Bessel beams have a set of concentric rings with constant size following the central spot, whereas the Hankel beams have cylindrical/conical characteristics. Thus, the HB vortex beams present in their shape a characteristic similar to these two separate beams, since they incorporate their structural elements. The normalized intensity projection and the phase map are observed in Fig. \ref{Fig_int_ph-1}, considering topological charge of $n = 1$.

\begin{figure}[H]
\centering
\includegraphics[width=12cm]{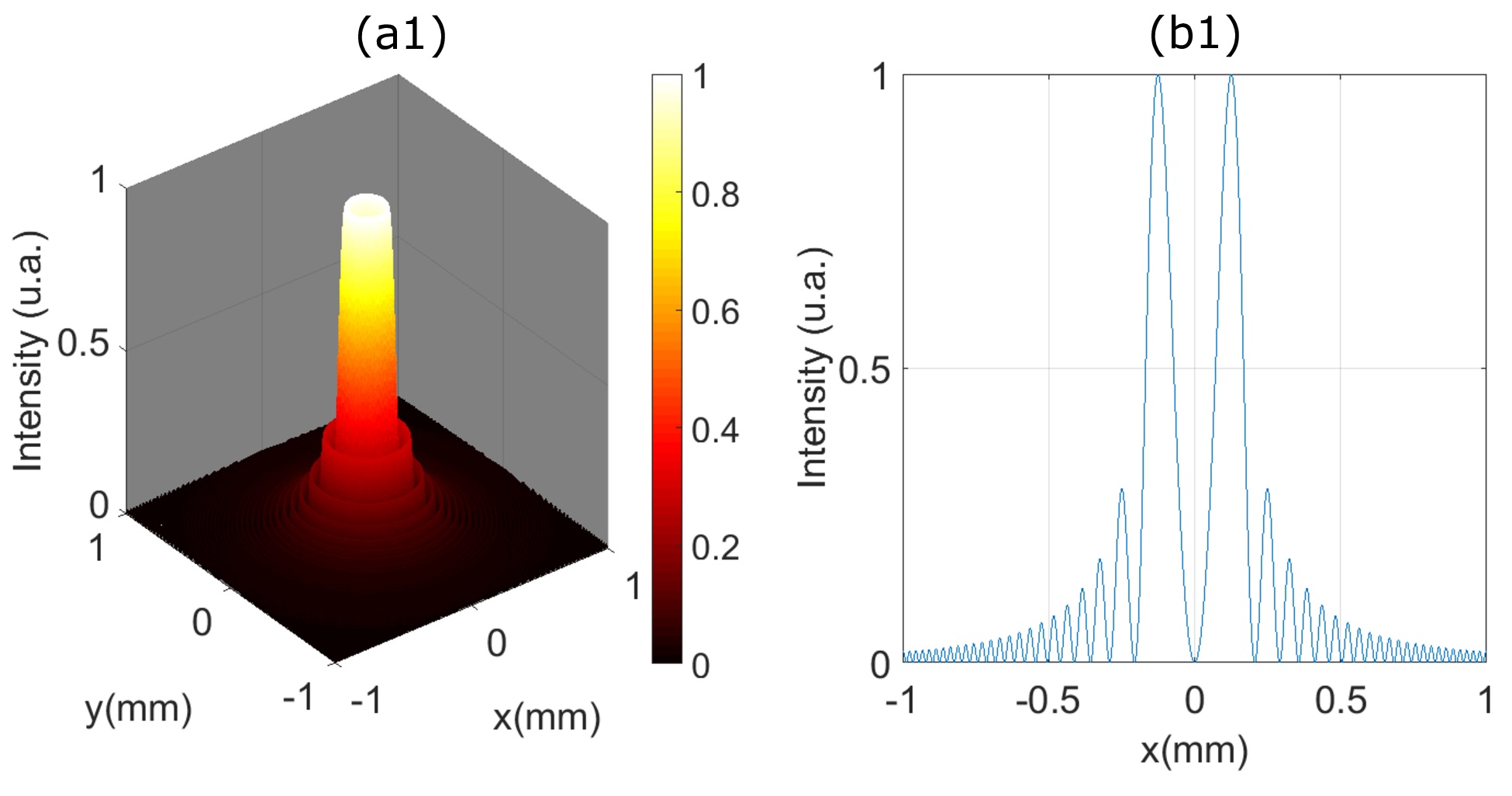}
\caption{Normalized 3D intensity profile (a1), and transverse intensity (b1), of integer HB vortex beams with $n = 1$.}
\label{Fig2HB_3D}
\end{figure}

In Fig. \ref{Fig2HB_3D}, it was observed how the transverse intensity profile of this beam behaves. Thus, both the 3D profile and the normalized intensity pattern were generated.

\begin{figure}[H]
\centering
\includegraphics[width=12cm]{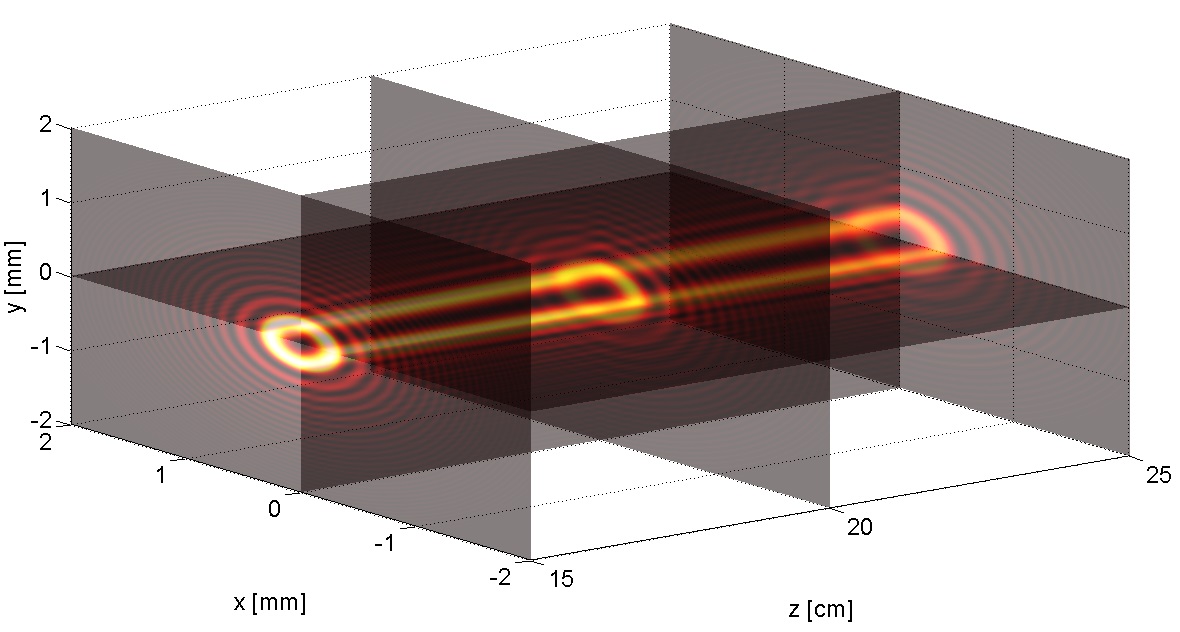}
\caption{Orthogonal propagation profile of the HB vortex beams, propagating from an initial arbitrary distance.}
\label{Fig3HB}
\end{figure}

Fig. \ref{Fig3HB} presents an orthogonal displacement profile of these beams from a determined point, arbitrarily defined. It's observed that beams with integer TC lose intensity and present a small divergence as they propagate along the z-axis.

\begin{figure}[H]
\centering
\includegraphics[width=13cm]{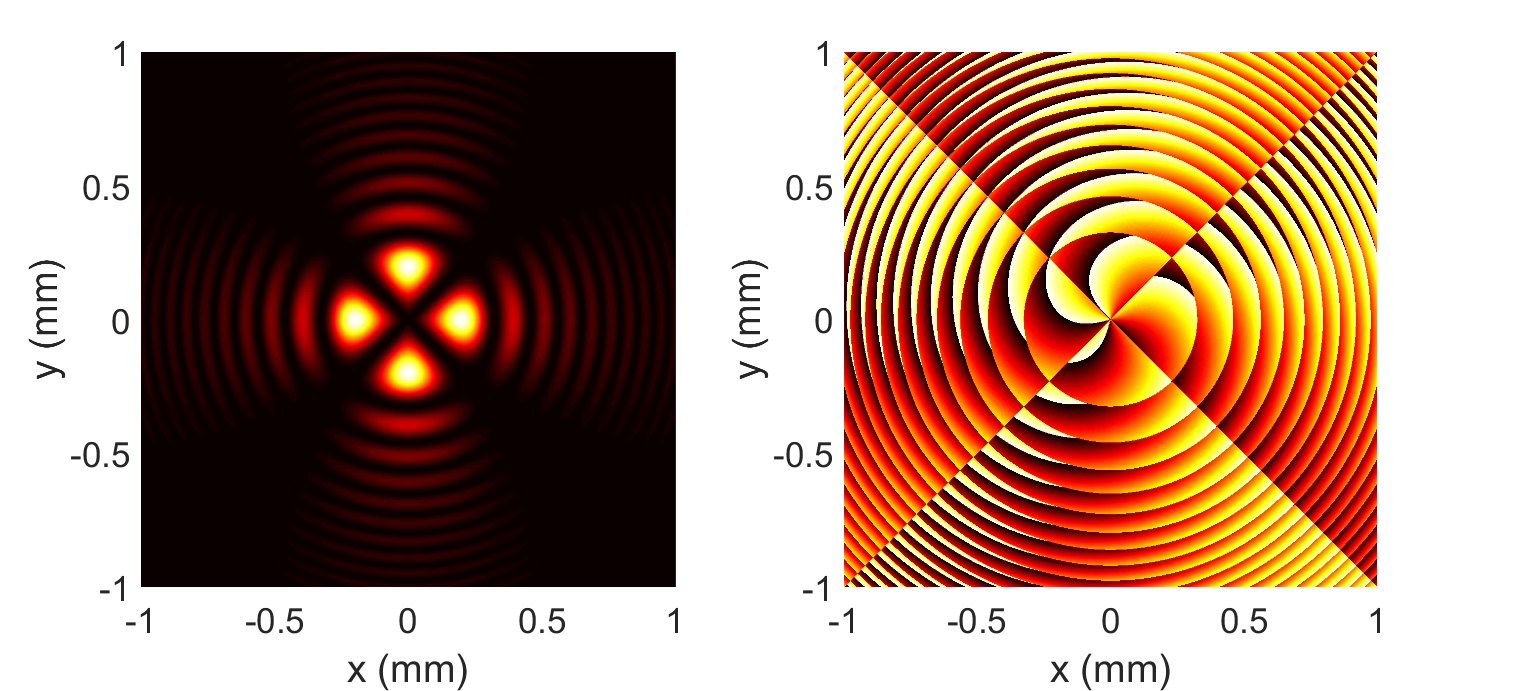}
\caption{Normalized intensity and phase map for the fractional HB vortex beams in 4 parts, $n = 3.9$.}
\label{Fig_int_ph-2}
\end{figure}

Initially, all studies and applications of these OV beams considered the values of the TC as an integer in which the helical phase had a displacement of $2 \pi n$ in the phase of the electromagnetic field. However, it was noted that TC can be a non-integer number, since it's possible to generate a phase jump that is not an integer multiple of $2 \pi n$ \cite{zhang2022review}. This beam with non-integer TC was called fractional vortex, and it was observed that it has a phase discontinuity and a low-intensity gap, because the intensity of the ring that forms the vortex is broken in the form of a dark radial opening, a sectioned or fractional ring \cite{zhang2022review, berry2004optical}.
\\
A fractional OV beam can be obtained by setting the TC of the beam to a fractional value, not an integer. This definition was widely spread, but it doesn't provide a proper solution to the Helmholtz equation \cite{zhang2022review, kotlyar2019calculation}. In this sense, a new approach was proposed which is based on the inverse Fourier transform \cite{zhang2022review}, and can be visualized in terms of the order $(n)$ of the beams, or also by the Fourier series, as described in \cite{berry2004optical}.
\\
The interest in these beams is due to their enormous field of application involving optical manipulation, because, while a beam with integer TC performs only one rotation in the ring of light, a beam with fractional TC carries an intensity distribution that can perform particle sorting or even particle orientation control much more accurately \cite{berry2004optical, gotte2008light}.
\\
The normalized intensity projection and the phase map are observed in Fig. \ref{Fig_int_ph-2}, considering a beam fractionated into 4 parts, $n = 3.9$.

\h {\em\bf 3. Experimental generation and analysis of the fractional HB vortex beams via holographic technique: Experiments and Results} 

\textbf{Computational holographic method}
Currently, computational methods allow the generation of the CGHs of optical beams using algorithms that characterize optical beams, particularly non-diffracting beams with OAM (optical vortex). The CGH record the hologram in a numerical and computational way \cite{benton2008, stein1992, yepes2017dynamic}, since the light beams are recorded and stored from computational algorithms, and can be reconstructed through a holographic optical system using a SLM \cite{hariharan1996, suarez2016photorefractive}.
\\
Through this computational technique, it's possible to generate experimental results with high quality and precision in the amplitude and phase reconstruction of structured light beams. In this work, we use equations \ref{eq:Sol_Eq_HB} and \ref{eq:hol_amp} to build the CGH of the HB vortex beams, and an SLM LETO that will perform the optical reconstruction of this hologram. The CGH is calculated by an amplitude transmission function, which varies the transmission or reflection coefficient of the medium \cite{suarez2020generation, gesualdi2023algoritmos2}:

\begin{equation}
H(x,y)=\frac{1}{2} \{ \beta(x,y)+\alpha(x,y)cos[\phi(x,y)-2\pi(\xi x + \eta y)] \},
\label{eq:hol_amp}
\end{equation}

where, $\alpha(x,y)$ and $\phi(x,y)$ describe the amplitude and phase of the complex field of the beam; $(\xi,\eta)$ the spatial frequency of the plane wave that was used as a reference and $\beta(x,y)=[1+\alpha^2(x,y)]/2$, is the function bias converted into a soft envelope of the complex amplitude $\alpha(x,y)$, used to reduce the influence of the central spectrum \cite{suarez2020generation, gesualdi2023algoritmos2,vieira2014modeling}.

On the other hand, the experimental analysis of the intensity and phase of these experimentally generated optical vortex can be done using interferometric techniques. Particularly, in this work we use the digital holography technique with the angular spectrum method to characterize the generated vortex. The computational reconstruction of the digital holograms of phase and intensity of the OV beams can be made by the angular spectrum method. The angular spectrum is defined as the Fourier Transform of the digital hologram. Through an inverse Fourier Transform, the field correspondent to this plane is calculated. This calculation results in a matrix of complex numbers and the phase of the object wave can be determined. This method is described in references \cite{yepes2019phase, suarez2020generation}.

\textbf{Holographic experimental setup}. SLMs are optoelectronic devices used in the computational holographic technique to experimentally generate optical beams and 3D images, allowing greater flexibility, efficiency and reliability of the results obtained in the holographic optical system \cite{cirino2011, gesualdi2023algoritmos, yepes2017dynamic, yepes2019phase, suarez2020generation, vieira2014modeling}.

\begin{figure}[H]
\centering
\includegraphics[width=14cm]{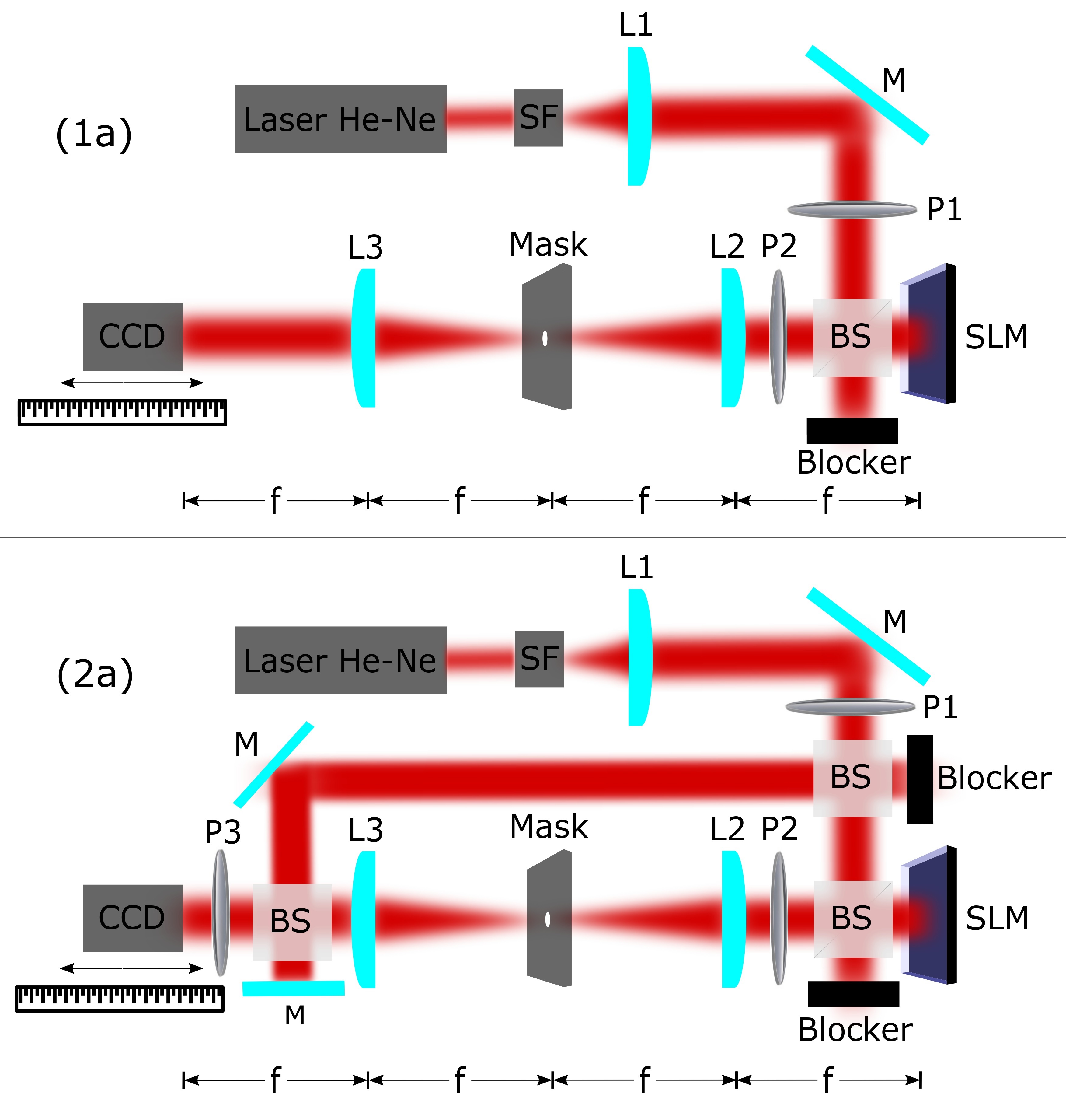}
\caption{Experimental optical system to holographic technique: (1a) for OV generation, and (2a) for OV generation and analysis; where, Laser He-Ne is a red laser, spatial filter (SF), lenses (L1, L2, L3), mirrors (M), polarizers (P1, P2, P3), beam splitter (BS) and spatial light modulator (SLM), Mask is a spatial mask, CCD is the camera.}
\label{Setup_exp}
\end{figure}

The experimental optical setup to generation of these OV from the CGHs of the HB vortex beams, integer and fractional, can be seen in Fig. \ref{Setup_exp}, where in Fig. \ref{Setup_exp} (1a) is the main configuration to generation of these OV beams.
\\
The light beam from the laser ($633~nm$ He-Ne and $35~mW$ power from Research Electro-Optics, Newport Corporation), is directed in the experiment to the spatial filter (SF), passing through the lens L1 where it is expanded and collimated. After being reflected by a mirror (M), this beam propagates towards the beam splitter (BS), so that part of the beam energy is directed towards the beam blocker and is lost, and the other part is directed and reflected on the surface of the SLM display that is aligned with the focus of lens L2.

Before reaching the BS, the beam find a polarizer P1 aligned at an angle of 0$^\circ$, and later, a second polarizer, at 90$^\circ$ in relation to the y-axis of the SLM. A LETO SLM, from Holoeye Photonics, with a screen resolution of 1920$\times$1080 and a pixel size of $6.4~\mu m$, was used, which is placed in the input plane (L2 lens focus). The beam reconstruction occurs due to the diffraction that occurs in the holographic lattice of the CGH in the SLM.

To select the different orders of diffraction that occur in the Fourier plane of the reconstructed CGH, a system, known as the 4f system, is used, consisting of two lenses, L2 and L3 (focal distance $150~mm$), in addition to a circular ID aperture (mask) that is used to selected the first order of diffraction (hologram information). After beam reconstruction, the 1280$\times$960 pixels monochrome USB CCD camera, model DMK 41BU02.H, from The ImagingSource, with a pixel size of $4.65~\mu m$, having a recording rate of up to 15 FPS (frames per second), which can move step by step along the beam propagation axis, captures the image that was generated.

Fig. \ref{Setup_exp} (2a) is the configuration for using digital holography (DH) to analyse the generation of these OV beams. In this case, a configuration of a Mach-Zehnder interferometer is added to the setup of Fig. \ref{Setup_exp} (1a), characteristic of the digital holography setup, for the generation of the HD of the OV. Subsequently, these HD are analyzed computationally using the angular spectrum method to obtain information on intensity and phase of the beam, as described in the references \cite{yepes2017dynamic, yepes2019phase, suarez2020generation}.

\textbf{Results}. For optical generation of the HB vortex beams, was used the experimental setup Fig. \ref{Setup_exp}-(1a). Initially, we build a CGH of the field described by Eq. \ref{eq:Sol_Eq_HB}  and the Eq. \ref{eq:hol_amp} adopting the carrier of frequencies $\eta=\xi=\Delta p/5$ for the plane wave of reference, where $\Delta p=1/\delta p$ is the bandwidth and $\delta p$ is the individual pixel size ($6.4~\mu m$).

\begin{figure}[H]
\centering
\includegraphics[width=6cm]{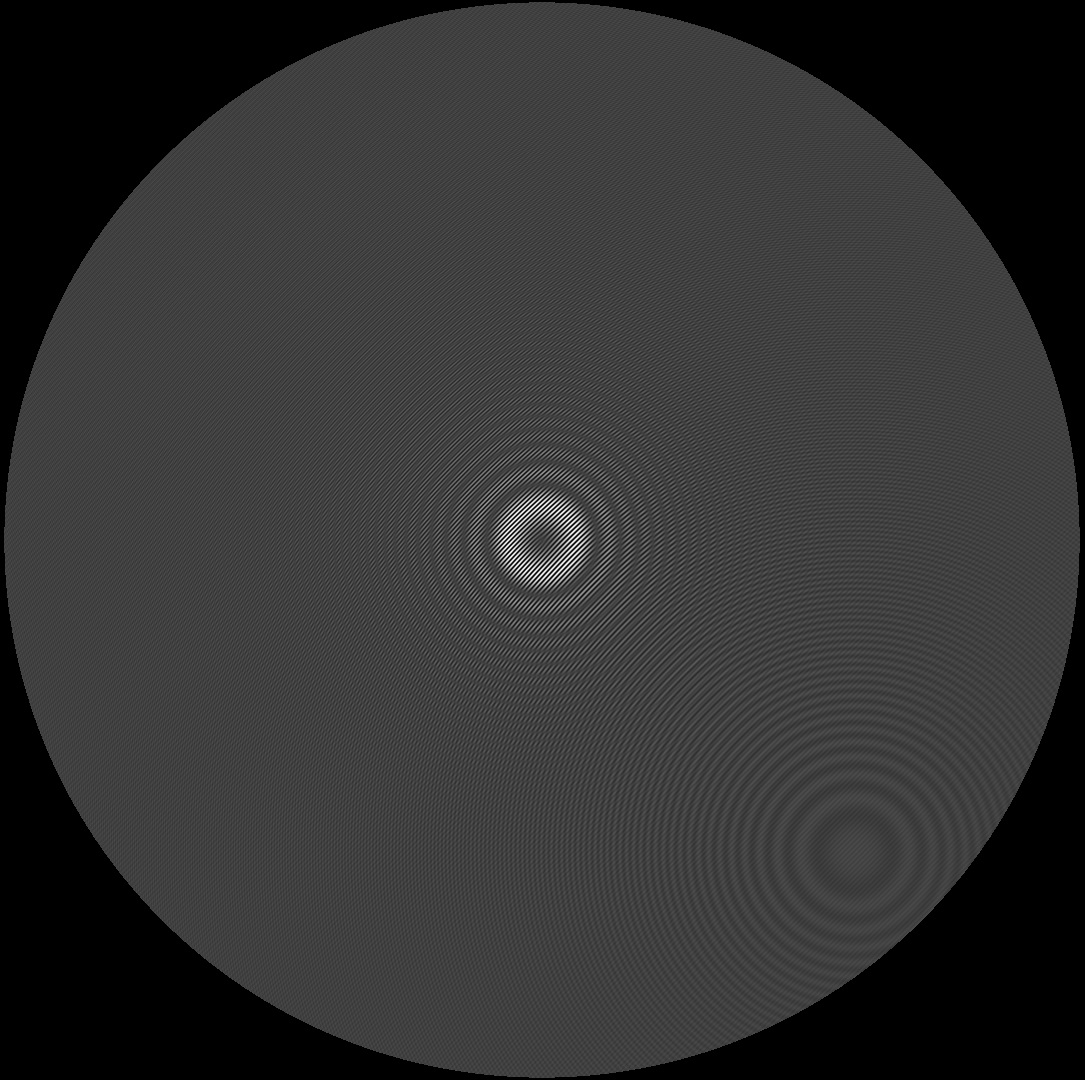}
\includegraphics[width=6cm]{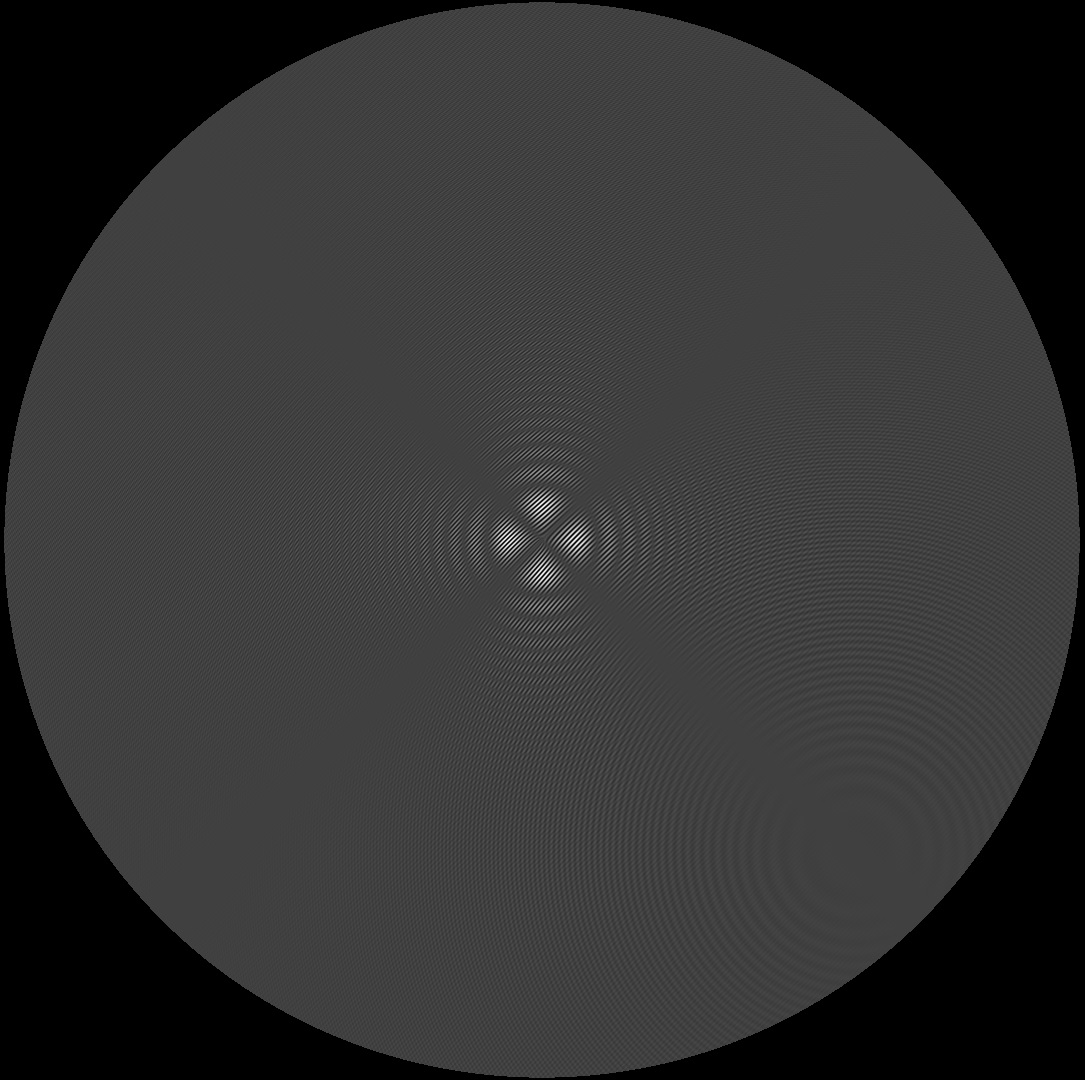}
\caption{CGH obtained from Hankel-Bessel vortex beams for integer ($n = 1$) and fractional (4 parts, $n = 3.9$) TC.}
\label{Fig_CGH_HB}
\end{figure}

The first results are presented in Fig. \ref{Fig_CGH_HB}, where it was possible to generate the CGH of the HB vortex beams, integer and fractional, and following all parameters on the SLM. To generate the fractionated beams, a TC divided into four parts ($n = 3.9$) was considered.

\begin{figure}[H]
\centering
\includegraphics[width=13cm]{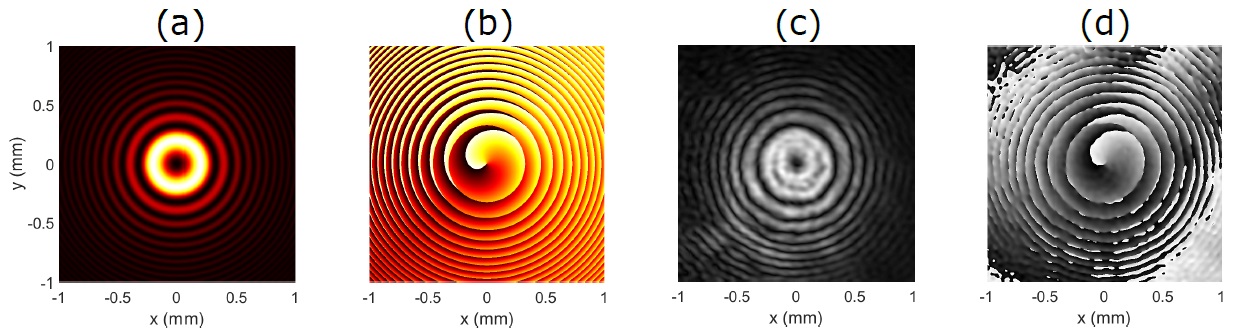}
\caption{Normalized intensity and phase map, theoretical $(a)-(b)$, and experimental $(c)-(d)$, for the HB vortex beams with $n = 1$.}
\label{Fig_theo_exp1}
\end{figure}

\textit{First case - HB vortex beams:} In this first case, considering the HB vortex beam with integer TC, $n = 1$, the intensity and phase map obtained through computer simulations $(a)-(b)$, and the first experimental results $(c)-(d)$ using the experimental setup described in Fig. \ref{Setup_exp}-(2a), where it's possible to compare these results through digital holography (DH). In Fig. \ref{Fig_theo_exp1}, the transverse distribution of the intensity and phase map results are presented.

The experimental results in Fig. \ref{Fig_theo_exp1}-(c) and \ref{Fig_theo_exp1}-(d), it's possible to observe, respectively: the intensity is zero and the formation of the main ring that makes up the HB vortex beams; and, the phase map of this vortex obtained experimentally by the DH technique, this is characterized by a helical structure that carries OAM and its magnitude depends on the TC. Furthermore, comparing the images, it’s possible to notice the similarity between the experimental and theoretical results. Then, the experimental results obtained through the digital holographic technique in Fig. \ref{Fig_theo_exp1}-(c) and \ref{Fig_theo_exp1}-(d), were in accordance with what was theoretically obtained and presented in Fig. \ref{Fig_theo_exp1}-(a) and \ref{Fig_theo_exp1}-(b). 

\begin{figure}[H]
\centering
\includegraphics[width=11cm]{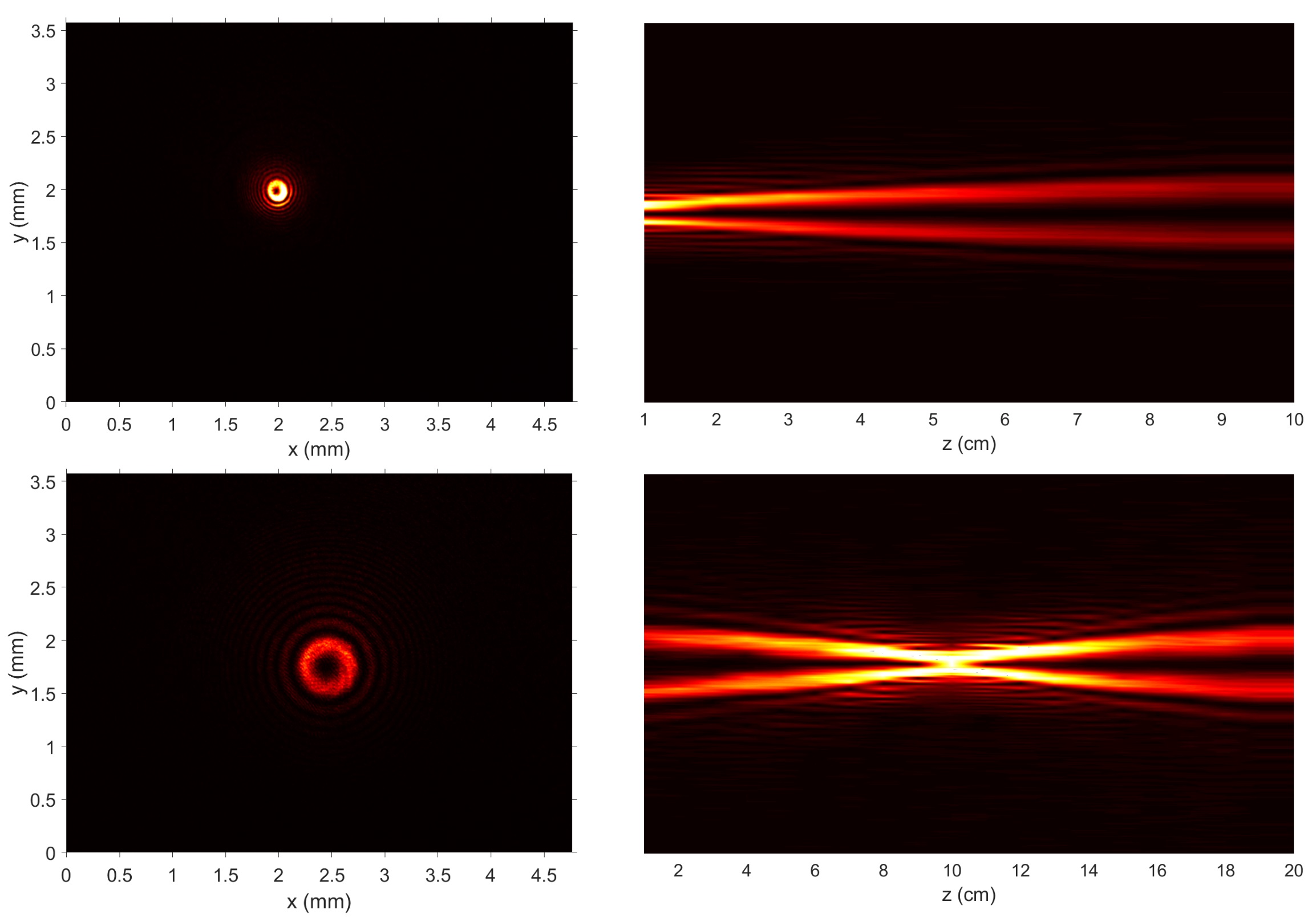}
\caption{Transverse pattern captured by CCD with integer TC ($n = 1$), and propagation of the HB vortex beam.}
\label{Fig_Transv}
\end{figure}

\begin{figure}[H]
\centering
\includegraphics[width=13cm]{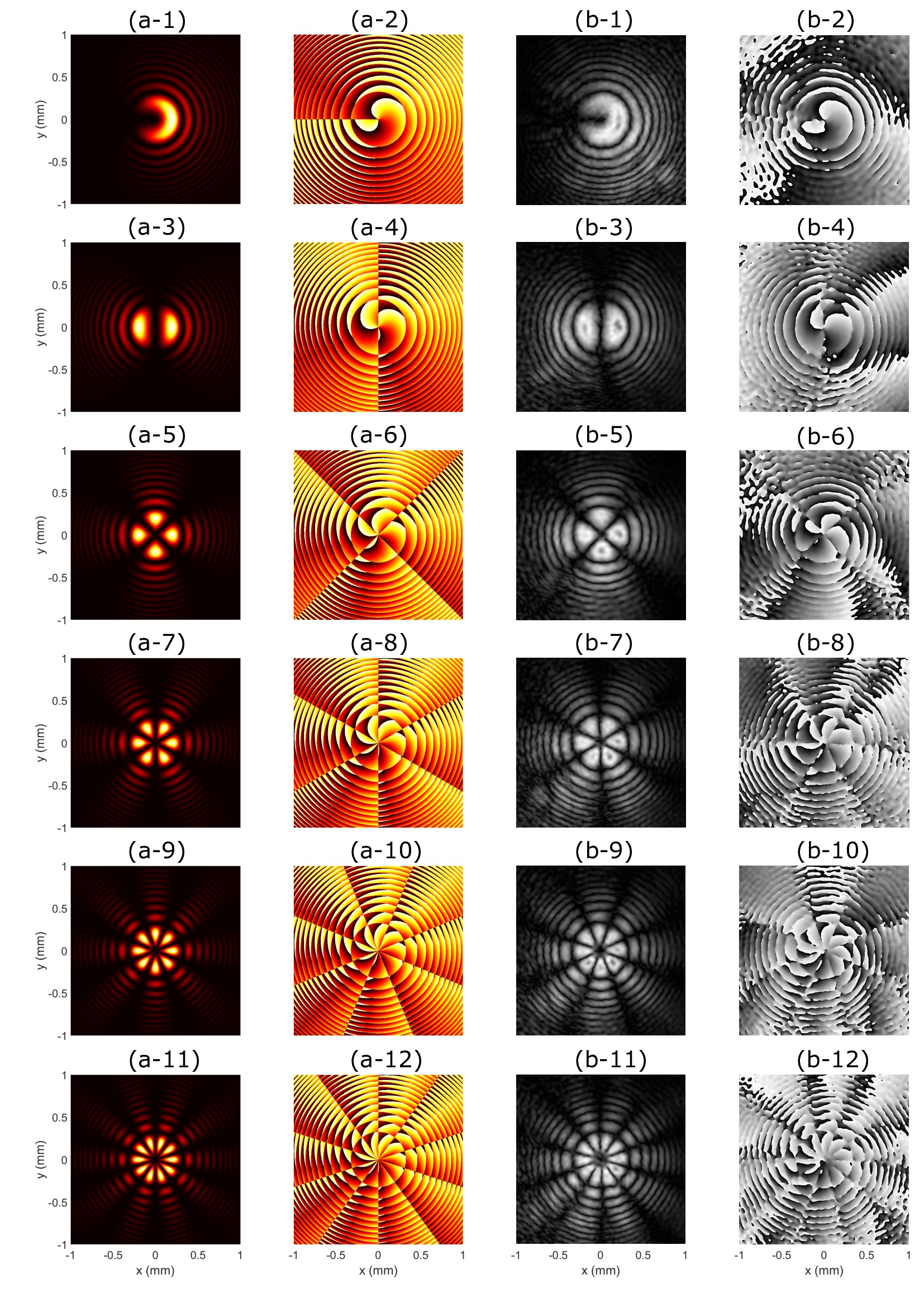}
\caption{Normalized intensity, theoretical $(a-M)$ and experimental $(b-M)$, where $M=1,3,5,7,9,11$ and phase map, theoretical $(a-N)$ and experimental $(b-N)$, where $N=2,4,6,8,10,12$, for the fractional HB vortex beams.} 
\label{Fig_theo_exp2}
\end{figure}

Fig. \ref{Fig_Transv} shows the experimental propagation of the HB vortex beam along the z-axis. It's possible to generate this vortex propagation once the hologram (CGH) has been generated through the process described, and its reconstruction is done by SLM. A CCD camera is used to observe and capture the different intensity distribution profiles along the z-axis. To build the transverse propagation profile of this vortex, the experimental apparatus shown in Fig. \ref{Setup_exp}-(1a) was used. It was possible to observe that the HB vortex beam propagation for approximate $10~cm$ along the z-axis, as can be seen in the first image of Fig. \ref{Fig_Transv}.
\\
In addition, the divergence theoretically described for this OV was observed in the laboratory, with integer TC, as can be seen in Fig. \ref{Fig_Transv} and corroborates what was observed and described in the literature. When this vortex propagates for large values of z, it's possible to observe the hyperbolic divergence described in the theory. Thus, this statement was confirmed experimentally when observing the propagation of the beam for a distance of $z = 20~cm$, as can be seen in the second image in Fig. \ref{Fig_Transv}.
\\
In this case, satisfactory results were obtained. This demonstrates that this technique is very useful to obtain results with precision and experimental compatibility not only in relation to optical beams, but also with OV.
\\
\textit{Second case - Fractional HB vortex beams:} For this second case, the fractionation of the HB vortex beams was considered. In this case, the fractional vortex were simulated and generated experimentally in the laboratory considering one ($n = 0.9$), two ($n = 1.9$), four ($n = 3.9$), six ($n = 5.9$), eight ($n = 7.9$) and ten ($n = 9.9$) fractional HB vortex beams.
\\
The theoretical transverse intensity and the phase map obtained through computer simulations $(a-1)$-$(a-12)$, and the experimental results $(b-1)$-$(b-12)$ are presented in Fig. \ref{Fig_theo_exp2}.
\\
The first fractional vortex ($n = 0.9$) can be observed in Fig. \ref{Fig_theo_exp2}, where the images $(a-1)$ and $(a-2)$ correspond to the simulated theoretical intensity and phase profile, respectively. In image $(b-1)$ we can see the experimental intensity profile captured by the CCD, in addition to observing a low-intensity gap, since the intensity of the ring that forms the vortex is broken in the form of a dark radial aperture. In the image $(b-2)$ we observe the profile of the experimental phase, and as already mentioned, there is a phase discontinuity characteristic of this OV fractionation. Fractionating the vortex into two parts ($n = 1.9$), we can observe that there is a break in intensity in the middle of the vortex, generating two points with greater intensity and a central gap that divides the beam. The images $(a-3)$ and $(a-5)$ correspond to the simulated theoretical intensity and phase profile, respectively. In image $(b-3)$ we can observe that the experimental intensity profile, as well as the phase profile, presented in image $(b-4)$, were in accordance with what was theoretically predicted. Analogously, we can observe the theoretical and experimental results of fractionation into 4 ($n = 3.9$), 6 ($n = 5.9$), 8 ($n = 7.9$) and 10 ($n = 9.9$) parts in figures $(a,b-5,6)$, $(a,b-7,8)$, $(a, b-9,10)$ and $(a,b-11,12)$, respectively.

\begin{figure}[H]
\centering
\includegraphics[width=10cm]{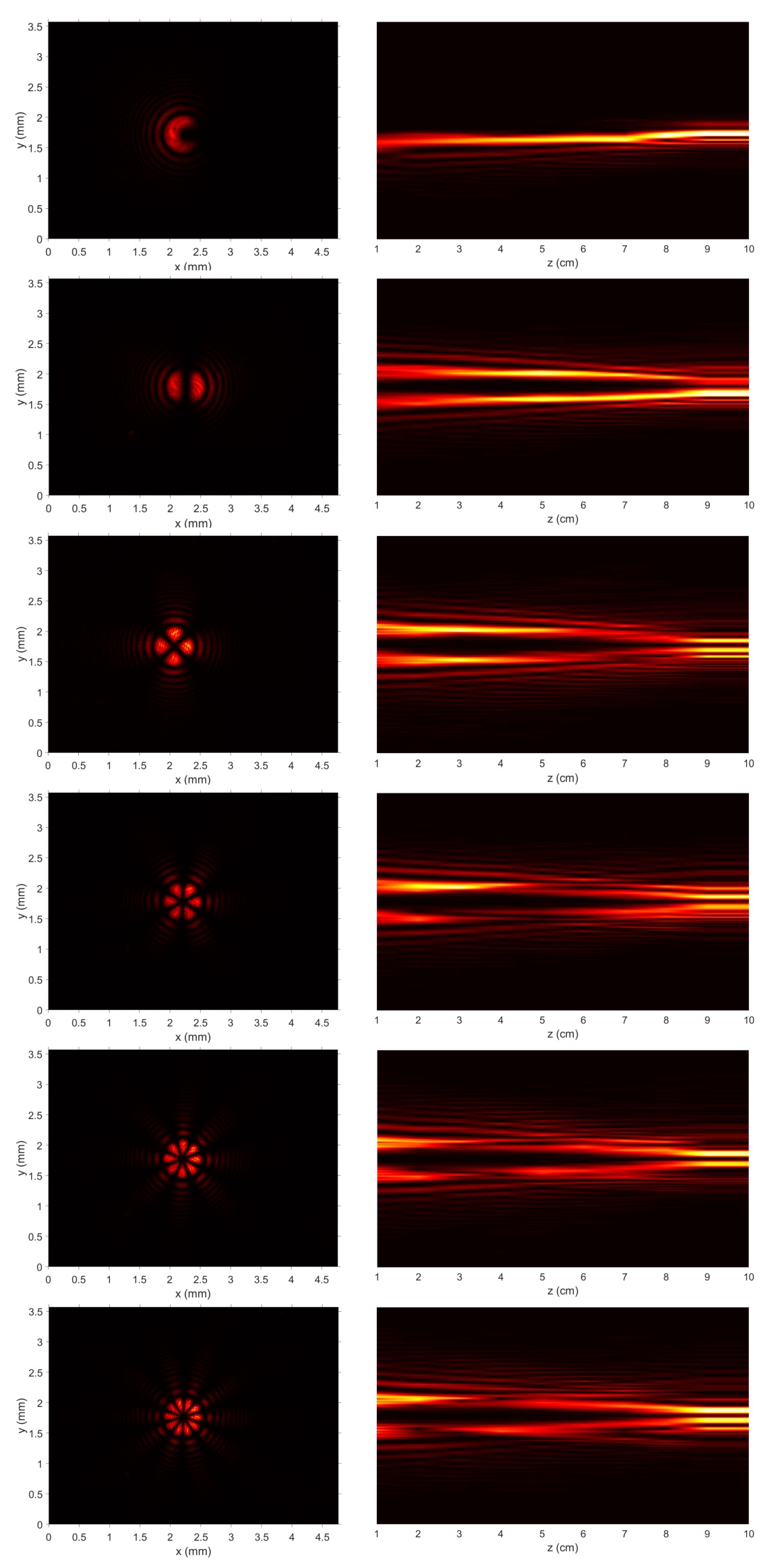}
\caption{Propagation profile along the z-axis for fractional HB vortex in 1, 2, 4, 6, 8 and 10 parts.}
\label{Fig_Transv2}
\end{figure}

Analyzing the propagation of fractional vortex along the z-axis, it was observed that their displacement is smaller compared to a vortex with integer TC. Similar to what was done for a TC of $n=1$, it was defined that the observation would occur at a distance of $z=10~cm$ for the fractional vortex. As the intensity and phase profiles of the fractional vortex, Frac-1, Frac-2, Frac-4, Frac-6, Frac-8 and Frac-10 were simulated, analyzed and described, these same vortex were also generated and their propagation profiles.

At this time, it was found that for any fractional OV, its propagation occurs up to approximately $8-9~cm$, moreover, there is no divergence observed as a integer TC. The holographic technique allows studying and analyzing the individual characteristics of each of these vortex just by varying the TC parameters.

The propagation of the first fractional vortex can be observed in Fig. \ref{Fig_Transv2}. Due to the fractionation in only one part, the transverse profile of this vortex presents only one component that describes the propagation of the beam. It can be seen that its displacement takes place at approximately $8~cm$, as previously mentioned. The second fractional vortex has two components that describe its propagation along the z-axis. Its displacement occurs at approximately $8-9~cm$, and from this point, there is no more relevant information about this beams. Analogously, we can observe the experimental propagation results of fractionation into 4, 6, 8 and 10 parts, respectively.

\h {\em\bf 4. Conclusions}  

In summary, this work presents the integer and fractional HB vortex beams and their experimental generation and analysis by a holographic setup along an optical reconstruction of dynamic holograms CGHs by using a SLM (and characterization of the phase distribution via DH). The experimental results, represented in the Fig. \ref{Fig_theo_exp1} and Fig. \ref{Fig_Transv} for the integer OVs, and in Fig. \ref{Fig_theo_exp2} and Fig. \ref{Fig_Transv2} for the fractionated OVs, presented interesting results and in agreement with what was foreseen in the literature. As the experimental results are in line with theoretical predictions, they open up exciting possibilities to generate many other potentially interesting optical vortex beams for scientific and technological applications in optical metrology, optics communications and, particularly, in optical manipulation for guiding and capturing particles. With this type of beams, several optical tweezers could be generated simultaneously, which would be suitable for capturing and rotating micro and nanoparticles, where the phase shape defines the particles direction of movement.

\h {\em Acknowledgments.} The authors acknowledge financial support from UFABC, CAPES, FAPESP (grant 16/19131-6) and CNPq (grant 302070/2017-6). And, the professors Erix A. M. Garces and Diogo Soga (IFUSP, São Paulo, Brazil) for their contribution with the instruments.

\end{document}